\documentclass[12pt,preprint]{aastex}
\newcommand{\etal}{{\it et al.\ }}

\newcommand{\eg}{{\it e.g.\ }}

\newcommand{\kms}{\mbox{$\rm km\,s^{-1}$}}

\newcommand{\lya}{\mbox{$\rm Ly\,\alpha$}}

\newcommand{\speco}[3]{\mbox{$\rm #1\,^{#2}\!#3^{o}$}}
\newcommand{\spece}[3]{\mbox{$\rm #1\,^{#2}\!#3^{e}$}}
\lefthead{Sigut, Pradhan, \& Nahar}
\righthead{Iron Emission from AGN}
 
\begin{document}

 

\title{Theoretical Fe\,{\sc i/ii/iii} Emission-Line
Strengths from Active Galactic Nuclei with Broad-Line Regions}

\author{T.\ A.\ A.\ Sigut}
\affil{Department of Physics and Astronomy,
{\it The\/} University of Western Ontario}
\authoraddr{London, Ontario Canada N6A 3K7}

\and

\author{Anil K. Pradhan \& Sultana N. Nahar}
\affil{Department of Astronomy, The Ohio State University}
\authoraddr{174 West 18th Avenue, Columbus, Ohio 43210-1106}

\begin{abstract}

We present theoretical iron emission line strengths for physical
conditions typical of Active Galactic Nuclei with Broad-Line Regions.
The non-local thermodynamic equilibrium (NLTE) models include a new and
extensive treatment of radiative transfer in the \ion{Fe}{3} ion,
complementing the \ion{Fe}{2} emission line strengths predicted in our
earlier works. We also briefly present preliminary results for the
\ion{Fe}{1} emission from AGN using a reduced atom model.  We can
satisfactorily reproduce the empirical UV \ion{Fe}{3} emission line
template of Vestergaard \& Wilkes (2001) for the prototypical
narrow-line Seyfert 1 galaxy I~Zw~1, both in terms of the general
\ion{Fe}{3} flux distribution and the relative strength of the
\ion{Fe}{3} and \ion{Fe}{2} emission. However, a number of  detailed
features are still not matched; the most prominent example is the
strongest single \ion{Fe}{3} feature observed in the I~Zw~1 spectrum,
UV47: it is predicted to be strong only in models suppressing Fe-H charge
exchange reactions.  We examine the role of variations in cloud
turbulent velocity and iron abundance and carry out Monte Carlo
simulations to demonstrate the effect of uncertainties in atomic data
on the computed spectra.

\end{abstract}


\section{Introduction}

The ultraviolet spectra of active galactic nuclei (AGN) with broad-line
regions (BLRs) exhibit a quasi-continuum of thousands of blended iron
emission lines, dominated by \ion{Fe}{2} (Wills \etal\ 1980, Wills,
Netzer, \& Wills 1980); numerous \ion{Fe}{2} transitions are also
present in the optical (Boroson \& Green 1992, V\'{e}ron-Cetty, Joly \&
V\'{e}ron 2004) and near-infrared (Rudy \etal 2000, Rudy \etal 2001,
Rodr\'{\i}guez-Ardila \etal 2002). Since their recognition as a
significant component of the BLR emission spectrum, these \ion{Fe}{2}
transitions have been the subject of intense observational scrutiny and
theoretical modelling (Netzer 1980; Kwan \& Krolik 1981; Netzer \& Wills
1983; Wills, Netzer \& Wills 1985; Elitzur \& Netzer 1985;
Collin-Souffrin \etal 1986; Penston 1987; Collin-Souffrin, Hameury \&
Joly 1988; Dumont \& Collin-Souffrin 1990; Netzer 1990; Sigut \&
Pradhan 1998; Verner \etal 1999; Sigut \& Pradhan 2003).  Reproducing
the strength of the \ion{Fe}{2} emission has been a considerable
challenge for traditional photoionized models for the BLR clouds;
typically the \ion{Fe}{2}/H$\beta$ flux ratio is observed to be
$\sim\,10$, ranging from $\sim\,2$ to near $\sim\,30$ in the case of
super-strong \ion{Fe}{2} emitters (Joly 1993; Graham, Clowes \&
Campusano 1996).

However \ion{Fe}{2} is not the only component of the low-ionization
emission spectrum of the BLR. Ultraviolet transitions of \ion{Fe}{3}
are also well established (Baldwin \etal 1996, Laor \etal 1997,
Vestergaard \& Wilkes 2001). Laor \etal (1997) identify a strong
feature near $\lambda\,2418\,$\AA\ in the spectrum of the prototypical
narrow-line Seyfert~1 (NLS1) galaxy I~Zw~1 as \ion{Fe}{3} multiplet
UV47.  Most recently, Vestergaard \& Wilkes (2001) provide a detailed
analysis of the UV \ion{Fe}{3} emission from the I~Zw~1 and use their
observations to empirically derive \ion{Fe}{2} {\it and\/} \ion{Fe}{3}
flux templates.

In our previous work (Sigut \& Pradhan 2003, hereafter SP03), we
constructed a detailed non-LTE radiative transfer model for
\ion{Fe}{2}, including 829 atomic levels, and predicted a large number
($\sim23,000$) of \ion{Fe}{2} emission line strengths for conditions
typical of BLRs of AGN.  In this paper, we extend our earlier work on
\ion{Fe}{2} to compute detailed non-LTE models for Fe\,{\sc i}-{\sc
iv}. We compare these predictions to the observed UV I~Zw~1 template.

%
%
%

\section{Atomic Data}

The four lowest ionization stages of iron, Fe\,{\sc i}-{\sc iv},
were explicitly included in the calculations (although \ion{Fe}{4} was
included as only a one-level atom).  A summary of the complete iron
model atom is given in Table~\ref{tab:iron_atom}. The final non-LTE
atomic model includes 944 energy levels and 14,962 radiative
bound-bound (rbb) transitions. With the exception of level energies,
the majority of radiative and collisional atomic data was computed
using methods developed under the Iron Project\footnote{Complete
references for the Iron Project are given at
www.astronomy.ohio-state.edu/$\sim$pradhan} (Hummer \etal 1993, IP.I). The
R-matrix method employed in the Iron Project (IP) ensures relatively
uniform accuracy for the computed data.  In the following sections, the
IP atomic data will also be identified according to the number of the paper
in the ongoing IP series published in {\it Astronomy and Astrophysics}
(\eg IP.I for the first paper in this series).

\subsection{\ion{Fe}{1}}

A minimal \ion{Fe}{1} atomic model was included in the calculation,
mainly to predict the ionization fraction of \ion{Fe}{1} in the various
BLR models and provide preliminary flux estimates.  While rare,
\ion{Fe}{1} fluxes have been tentatively identified for a few
\ion{Fe}{2}-strong quasars (Boroson \& Green 1992, Kwan \etal 1995).
The \ion{Fe}{1} atomic model consisted of 77 fine-structure levels (see
Figure~\ref{fig:fe1_grotrian}) comprising 19 low-energy LS levels from
the triplet, quintet, and septet spin symmetries.  Energy levels and
Einstein spontaneous transition probabilities ($A_{ji}$ values) for the
185 transitions were taken from the NIST
compilation\footnote{http://physics.nist.gov/PhysRefData/contents-atomic.html}.
Photoionization cross sections were adopted from the
\underline{R}-matrix calculations of Bautista \& Pradhan (1997,
IP.XX). Low-temperature, IP effective collision strengths for electron
impact excitation of the first 10 meta-stable \ion{Fe}{1} levels have
been computed by Pelan \& Berrington (1997, IP.XXI). All remaining
collision strengths were estimated with the effective Gaunt factor
approximation of van Regemorter (1962).

\subsection{\ion{Fe}{2}}

Atomic data for \ion{Fe}{2} has been extensively discussed by Sigut \&
Pradhan (1998, 2003). The IP data for collisions strengths for electron
impact excitation were computed by Zhang \& Pradhan (1995, IP.VI), for
radiative transition probabilities, by Nahar (1995, IP.VII), and for
photoionization cross sections, by Nahar \& Pradhan (1994).  We have
adopted the smaller, 256 fine-structure level atom of Sigut \& Pradhan
(1998) for use in this calculation. The smaller number of \ion{Fe}{2}
levels allows a maximal number of \ion{Fe}{3} energy levels to be
simultaneously included.  However, this smaller \ion{Fe}{2} atom is
still sufficiently large for realistic estimates of the flux and the
inclusion of \lya\ fluorescent excitation.

Among the atomic data for \ion{Fe}{2}, we call particular attention to
the $\rm Fe^{2+}+H \Leftrightarrow Fe^{+}+H^{+}$ charge-exchange
reaction rates of Neufeld \& Dalgarno (1987). These large rates,
computed in the Landau-Zener approximation, play an important role in
the \ion{Fe}{2}-\ion{Fe}{3} ionization balance. However, their accuracy
and efficacy in determining emission line strengths can be a source
of uncertainly in the present calculations, as discussed later.

\subsection{\ion{Fe}{3}}

\ion{Fe}{3} was represented by 581 fine-structure energy levels
(see Figure~\ref{fig:fe3_grotrian}) which includes all \ion{Fe}{3} levels
with known energies (see the NIST compilation; also Nahar \& Pradhan
1996).  We have used four principal sources to compile the required
$A_{ji}$ values for \ion{Fe}{3} transitions: the NIST
critically-evaluated database; the \underline{R}-matrix results of
Nahar \& Pradhan (1996, IP.XVII) for dipole-allowed ($\Delta\,S=0$)
transitions; the Hartree-Fock calculations of Ekberg (1993), which also
treats inter-system ($\Delta\,S\ne\,0$) transitions, and the extensive
semi-empirical calculations of Kurucz (1992).

Photoionization cross sections were adopted from Nahar \& Pradhan
(1996, IP.XVII).  Collision strengths for electron impact excitation
were adopted from Zhang (1996, IP.XVIII). All remaining collision
strengths were estimated with the Gaunt factor approximation. Charge
exchange reactions with \ion{Fe}{4} were also included following
Kingdon \& Ferland (1996).

\section{Calculations}

The iron line fluxes were calculated using the procedure of SP03.
Briefly, a background temperature and density structure for a single
BLR cloud of a given ionization parameter and total particle density
was computed with {\sc cloudy} (Ferland, 1992).  The shape of the
photoionizing continuum was taken to be that of Mathews \& Ferland
(1987). Given this fixed, one-dimensional model, the coupled equations
of radiative transfer and statistical equilibrium were solved for a
self-consistent set of iron level populations and line fluxes.  A
complete discussion of the system of equations solved, including the
treatment of fluorescent excitation by Lyman-$\alpha$ and
Lyman-$\beta$, and the implementation details, are given in SP03. We
note that our treatment of fluorescent excitation involves first the
construction the full frequency-dependent, PRD (partial-redistribution)
source functions for Lyman-$\alpha$ and Lyman-$\beta$, and then their
inclusion in the total monochromatic source functions used in the
radiative transfer solutions for all overlapping iron rbb transitions.

In the current work, each set of iron fluxes is specified by four model
parameters representing the BLR cloud: the ionization parameter, the
total particle density, the (internal) cloud turbulent
velocity\footnote {Turbulent velocity is used to refer to what is
commonly denoted {\it microturbulence\/} by stellar astronomers; this
velocity represents the width of a Gaussian distribution of turbulent motions
small in scale compared to unit optical depth. These motions act to
broaden the atomic absorption profile and thus can alter the
radiative transfer solution in each line.}, and the iron abundance
relative to hydrogen. Table 2 identifies the different BLR models
constructed in this work and the associated ionization parameters and
particle densities.

\section{Iron Ionization Balance}
\label{sec:ion}

The ionization balance among the four lowest ionization stages of iron
was self-consistently calculated simultaneously with the level
populations. Expression for the total and level-specific photo and
collisional ionization-recombination rates are given in Paper~I.  A
small correction was applied at each depth for the fraction of iron
more highly ionized than \ion{Fe}{4}, with this correction being
estimated by {\sc cloudy}.

Our self-consistent procedure is different from Verner \etal (1999) who
first solve for the iron ionization balance in isolation, balancing
photoionization from the ground states against effective recombination
rates, and then solving for the individual (\ion{Fe}{2}) level
populations using a fixed ionization structure.  While this
approximation is typical of nebular work where the population in the
excited states is very small, densities in the BLR may approach $\sim
10^{+9}-10^{+11}$ particles per cm$^{3}$, high enough to warrant a
detailed treatment of photoionization from excited states in the
calculation of the ionization balance. Another advantage of our
approach is that no {\it a priori\/} assumptions are made as to the
optical depths in the radiative transitions forming the recombination
cascade (assumptions usually expressed as ``Case A'' or ``Case'' B,
following Osterbrock, 1989, for example); a self-consistent solution is
automatically obtained.

Figure~\ref{fig:ion_bal_vt10_Fep0} shows the \ion{Fe}{1}-\ion{Fe}{4}
ionization balance for all the models of Table~\ref{tab:BLR} assuming a
solar iron abundance and turbulent velocity of 10\,\kms (at
$10^{4}\;$K, the thermal widths of the iron lines are $\sim 2\;$\kms).
At large optical depths in the Lyman continuum, \ion{Fe}{2} is {\it
always\/} the dominant iron species. However, closer to the illuminated
face, the ionization parameter plays a strong role; for the lowest
ionization parameter considered, \ion{Fe}{3} dominates in the outer
layers, while at the higher ionization parameters, \ion{Fe}{4} is the
dominant species, with \ion{Fe}{3} confined to a narrow zone at
intermediate Lyman continuum optical depths. Also shown in
Figure~\ref{fig:ion_bal_vt10_Fep0} is the hydrogen ionization balance
as predicted by {\sc cloudy}.  The ionization of hydrogen has a strong
influence on the iron ionization balance not only through the obvious
effect of its opacity on the local ionizing radiation field, but also
through strong charge-exchange reactions, $\rm Fe^{2+}+H
\Leftrightarrow Fe^{+}+H^{+}$.

Figure~\ref{fig:ion_bal_vt10_Fep0_ncex} shows the predicted ionization
fractions in the same BLR models as Figure~\ref{fig:ion_bal_vt10_Fep0}
except neglecting Fe-H charge-transfer reactions. There are
significant differences, most notably for the higher ionization
parameters at the lower particle density considered, models u20h96 and
u13h96. The absence of charge exchange reactions in these models leads
to \ion{Fe}{3}, and not \ion{Fe}{2}, as the dominant iron ionization
stage throughout the cloud model. The influence of the charge-exchange
reactions on the predicted iron fluxes will be discussed in a later
section; such models {\it may\/} be partially supported by observation
of I~Zw~1 as they are better able to account for an intense feature
near $\lambda\,2418\;$\AA\ identified by Laor \etal (1997) as
\ion{Fe}{3} UV47. We note that the charge exchange ionization and
recombination rates for $\rm Fe^{2+}+H \Leftrightarrow Fe^{+}+H^{+}$
predicted by Neufeld \& Dalgarno (1987) were obtained in the rather
crude Landau-Zener approximation (see Flower 1990, also Kingdon \&
Ferland 1996) and thus it is not inconceivable that this rate may be
significantly overestimated.

Finally we note that the predicted fraction of \ion{Fe}{1} in all cases,
including the lowest ionization parameter considered, is very small, $\ll
10^{-3}$. The question of weather or not such a small neutral fraction
can account for the tentatively identified \ion{Fe}{1} features in AGN
spectra will be addressed in a later section.

\section{Predicted \ion{Fe}{3} Fluxes}

Table~\ref{tab:iron_per_fluxes} gives the total iron fluxes in the
wavelength interval of 1500-5000\,\AA\ for the single-zone BLR cloud
models of Table~\ref{tab:BLR}. The effect of increasing the cloud
turbulent velocity from 10 to 50~\kms\ and of tripling the iron
abundance relative to hydrogen are also given.  The total flux is
broken down into the percentage contributions of each of the iron
ions.  The higher density models with the largest ionization
parameters, models u20h11 and u13h11, predict the largest contribution
of the flux arising from \ion{Fe}{3}, approaching $\sim 35$\% in model
u13h11.  As also seen from the table, the percentage contribution of
\ion{Fe}{3} tends to decrease with increasing turbulent velocity but
increases with an enhanced iron abundance; the u13h11 model with a
threefold enhancement of the iron abundance comes close to having an
equal flux split between \ion{Fe}{2} and \ion{Fe}{3}. It should be
borne in mind that these comparisons are with the significantly smaller
atom \ion{Fe}{2} of Sigut \& Pradhan (1998) to allow the simultaneous
treatment of the current large \ion{Fe}{3} atom.

Figures~\ref{fig:flux_tagn20_fe123} and \ref{fig:flux_tagn70_fe123}
show the wavelength distributions of the emitted iron line fluxes for
the u20h96 and u13h11 models, respectively, in the case of a solar iron
abundance and 10\,km/s of turbulent velocity. The lower wavelength
limit has been extended down from the 1500\,\AA\ of
Table~\ref{tab:iron_per_fluxes} to 500\,\AA\ in order to show the
complete \ion{Fe}{3} spectrum. The separate panels in these
plots show the total flux as well as the contributions from the
individual iron ions.  There is a striking difference in the
\ion{Fe}{3} fluxes predicted for these two models. For u20h96, the
strongest \ion{Fe}{3} emission occurs at wavelengths near 1900\,\AA.
These transitions represent decays from the lowest odd-parity
\ion{Fe}{3} level, \speco{z}{7}{P}, to \spece{a}{7}{S} and are the only
possible $\Delta\,S=0$ transitions (see
Figure~\ref{fig:fe3_grotrian}).  Inter-system ($\Delta\,S\ne0$)
transitions from \speco{z}{7}{P} to \spece{a}{5}{D}, the \ion{Fe}{3}
ground state, giving lines near $\lambda\,1220\,$\AA, and to
\spece{a}{S}{5}, giving lines near $\lambda\,2400\,$\AA\ (multiplet
UV47), are also present but weak. The multiplet UV47 transitions
identified as the strongest \ion{Fe}{3} features in the spectrum of
I~Zw~1 by Laor \etal (1997) will be further discussed in
Section~\ref{sec:2400}. Decays from the only other low-lying odd-parity
level, \speco{z}{5}{P}, about $0.8\,$eV above \speco{z}{7}{P}, are not
prominent in the predicted spectrum for u20h96. A number of
$\Delta\,S=0$ decay paths from this level are possible, including the
\ion{Fe}{3} ground state \spece{a}{5}{D} resulting in transitions near
$\lambda\,1120\,$\AA, as well as decays to \spece{a}{5}{P} and
\spece{b}{5}{D} giving rise to transitions with $\lambda>4000\,$\AA.

The higher ionization parameter, higher density model u13h11 presents a
radically different picture. Now, the region around
1000\,\AA\ dominates the \ion{Fe}{3} emission. This region is filled
with decays from the numerous odd-parity levels near $\sim15\,$eV to
the lowest even parity levels ($<5\,$eV) of each spin system. The
spectral region around $\lambda\,2000\,$\AA\ is now filled with strong
\ion{Fe}{3} emission arising principally from the decay of odd parity
levels to the numerous even parity levels near $\sim10\,$eV. As noted
in Table~\ref{tab:iron_per_fluxes}, a significant fraction of the total
iron flux comes out in the form of \ion{Fe}{3} transitions.

In both of these models, fluorescent excitation by Lyman-$\alpha$ is
unimportant due to the absence of significant line-overlap with any
\ion{Fe}{3} transitions.  However, fluorescent excitation by
Lyman-$\beta$ offers more potential.  There are transitions from
several low-lying, even-parity triplet states within a few
{\AA}ngstroms of Lyman-$\beta$ (at $1025.72\,$\AA):
\spece{a}{3}{P}-\speco{z}{5}{S} at $1025.71\,$\AA\ ($f_{ij}=0.0022$),
\spece{a}{3}{D}-\speco{y}{3}{P} at $1024.11\,$\AA\ ($f_{ij}=0.023$),
and \spece{a}{3}{D}-\speco{y}{3}{P} at $1026.79$,
$1026.88\,$\AA\ ($f_{ij}=0.026$,$f_{ij}=0.0011$). Re-performing the
u20h96 model without Lyman-$\beta$ fluorescent excitation resulted in a
reduction in the total predicted \ion{Fe}{3} flux of only about 12\%;
the reduction in the case of u13h11 from the absence of Lyman-$\beta$
fluorescent excitation was negligible.

\section{Predicted \ion{Fe}{1} Fluxes}

According to Table~\ref{tab:iron_per_fluxes}, $\sim0.1-5$\% of the
iron line flux comes out in transitions of \ion{Fe}{1}. In
Figure~\ref{fig:flux_tagn70_fe123}, representing model u13h11, the
predicted emission due to \ion{Fe}{1} in the $2700$-$4500\;$\AA\ region
is, although quite weak, not entirely negligible. This is illustrated
in Figure~\ref{fig:iras_fe1}; the strongest transitions present in this
figure are just those expected from the \ion{Fe}{1} Grotrian diagram of
Figure~\ref{fig:fe1_grotrian}, namely decays from low-lying odd parity
levels between $\sim3.5$ to $4.5\,$eV to the three lowest even parity
states, the \spece{a}{5}{D} ground state, \spece{a}{5}{F}, and
\spece{a}{3}{F}. The strongest predicted multiplet is 23,
\speco{z}{5}{G} - \spece{a}{5}{F}, giving lines near
$\lambda\,3600\,$\AA. Such multiplets of \ion{Fe}{1} have been
tentatively identified in PHL~1092 by Bergeron \& Knuth (1980) and Kwan
\etal (1995) and in IRAS~07598+6508 by Kwan \etal\ Both objects are
\ion{Fe}{2}-strong quasars. The current calculations tend to support
these identifications and the occurrence of \ion{Fe}{1} transitions in
some AGN spectra. Comparing our predictions to the IRAS 07598+6508
spectrum given by Kwan \etal (their Figure~1), we note that the
observed \ion{Fe}{1} flux relative to the nearby \ion{Fe}{2} flux
between $\lambda\,3600-3800\,$\AA\ is much larger than predicted by our
models, even when considering the effect of the tentative \ion{Ti}{1}
blends identified by Kwan \etal\  However, it is likely that our models
underestimate the \ion{Fe}{1} flux; the very limited \ion{Fe}{1} atomic
model employed artificially suppresses recombination by omitting energy
levels within $\sim3\;$eV of the continuum. This acts both to reduce
the recombination contribution the line fluxes and to lower the
\ion{Fe}{1} ionization fraction which also leads to weaker lines.  We
plan to extend our treatment of the \ion{Fe}{1} atomic model in the
future to provide more realistic flux estimates to compare with these
observations.

\section{Dependence of Iron Flux on BLR Cloud Parameters}

Figure~\ref{fig:fe123_parplot} summarizes the dependence of the
\ion{Fe}{1}, \ion{Fe}{2}, and \ion{Fe}{3} fluxes on the various model
parameters. The trend of increasing iron line fluxes with both
ionization parameter and particle density is evident, as well as the
maximization of the \ion{Fe}{3} fluxes for the higher ionization
parameter models. The figure also shows increased flux in \ion{Fe}{2}
and \ion{Fe}{3} for either an increased cloud turbulent velocity or an
increased iron abundance. Interestingly, however, the predicted
\ion{Fe}{1} flux does not depend on the internal cloud turbulent
velocity. The lines of \ion{Fe}{1} are weak and unsaturated so are
insensitive to the turbulent velocity, contrary to the lines of the
other ionization stages. From this figure it might seem that the
internal cloud turbulent velocity might be deducible from the spectrum
of an object exhibiting lines of both \ion{Fe}{1} and \ion{Fe}{2}. In
fact, this is the classic technique in stellar atmospheres to determine
the turbulent velocity dispersion (often dubbed
microturbulence)--forcing weak and strong lines to be reproduced in
strength by the same abundance.  It might be worthwhile to attempt this
in AGN; however, the situation is much more complex as the entire BLR
spectrum is likely not formed within a single cloud or ensemble of
identical clouds (Baldwin \etal 1995); the question would arise as
to whether clouds of different ionization parameters and particle
densities would have similar turbulent velocities. Nevertheless it
would be still worthwhile to attempt this investigation. Knowledge of
the intrinsic width of an emission line for individual clouds in the
BLR has important implications on the uncertain nature of the BLR
clouds themselves.  Narrow, near thermal widths of
$\sim10\;$\kms\ imply large numbers of BLR clouds in oder to
reproduce the smooth observed line profiles. Much larger intrinsic
widths would require fewer clouds to produce smooth profiles, perhaps
allowing stellar models for the BLR clouds to become viable (Peterson,
Pogge \& Wanders, 1999).

\section{Comparison to I~Zw~1 \ion{Fe}{2}-\ion{Fe}{3} UV Template}

Vestergaard \& Wilkes (2001) provide an empirical
\ion{Fe}{2}-\ion{Fe}{3} UV template
($1250\,\le\,\lambda\,\le\,3080\,$\AA) derived from I~Zw~1, the
prototypical narrow-line Seyfert 1 galaxy. One of the main results of
this work is the careful documentation of extensive \ion{Fe}{3}
emission from I~Zw~1, building on the identifications of 
Laor \etal (1997). Figure~\ref{fig:panel_1zw1} compares four of our
basic models with a solar iron abundance and minimal turbulent velocity
(10$\;$\kms) with the I~Zw~1 template. In these comparisons, the
calculations have been broadened by convolution with a Gaussian to
$500\,$\kms (FWHM), and the template has been normalized to the median model
flux in the region of strongest \ion{Fe}{3} flux included in the template
wavelength region, $1800$-$2000\;$\AA.
Only our highest-density, highest ionization parameter model, u13h11,
correctly reproduces the overall level of the \ion{Fe}{2} UV
emission line strength in the $2000$-$2500\;$\AA\ region. Thus we find that
there is, in principle, no problem in accounting for the basic level of
\ion{Fe}{3} flux from I~Zw~1, even with a crude, single-zone model.

While inspection of Figure~\ref{fig:panel_1zw1} clearly reveals a
strong correlation between the model and the empirical template, there
are several disagreements in detail. In the principle \ion{Fe}{3}
wavelength region, $1800$-$2000\,$\AA, the relative strengths of
individual features are not correctly reproduced. Most notable,
however, is the failure of the basic model to correctly reproduce the
strength of the \ion{Fe}{3} feature near
$\lambda\,2418\,$\AA\ identified by Laor \etal (1995).  This strong
feature is associated with \ion{Fe}{3} multiplet UV47, the inter-system
transition \speco{z}{7}{P}-\spece{a}{5}{S}, giving (identified)
features near $\lambda\,2418.58\,$\AA\ ($J=3-2$, $f_{ij}=0.0027$) and
$\lambda\,2438.18\,$\AA\ ($J=2-2$, $f_{ij}=0.0011$).  Finally, some of
the disagreement in the $2200$-$2600\,$\AA\ region can be traced to
our use of a limited \ion{Fe}{2} model atom.

To assess the seriousness of these discrepancies, we have performed
additional calculations along two fronts: first, we considered
predictions of models in which the highly-uncertain Fe-H charge-exchange
reactions were omitted. As noted in
section~\ref{sec:ion}, omission of these rates can substantially alter
the iron ionization balance for certain models; secondly, we have
performed a series of Monte Carlo simulations to assess the effect of
errors in the basic atomic data have in the predicted \ion{Fe}{3} line
fluxes. We shall first discuss models omitted charge-exchange
reactions.

\subsection{Charge-Exchange and \ion{Fe}{3} UV47 in I~Zw~1}
\label{sec:2400}

The important role played by the rather uncertain Fe-H charge
exchange reaction rates motivated us to compute a series of models in
which these rates were omitted. The predicted iron ionization balances
have already been discussed in section~\ref{sec:ion} with the results
shown in Figure~\ref{fig:ion_bal_vt10_Fep0_ncex}.  The predicted flux
of model u20h96 without charge-exchange reactions is shown in
Figure~\ref{fig:1zw1_tagn20_ncex} and these predictions may be compared
with Figure~\ref{fig:flux_tagn20_fe123}. As expected by the change in
the ionization balance shown in Figures~\ref{fig:ion_bal_vt10_Fep0}
and \ref{fig:ion_bal_vt10_Fep0_ncex}, the model without
charge-exchange reactions has a lower \ion{Fe}{2} flux and a higher
\ion{Fe}{3} flux. It is still the case that normalizing the predicted
iron fluxes to the observed template in the interval
$1800$-$2000\,$\AA\ results in an over-prediction of the
\ion{Fe}{2} flux, but the over-prediction is now less. Now conspicuous
in the model omitting charge-exchange reactions is the presence of
strong lines of \ion{Fe}{3} multiplet UV47.  However, this is not
translated into a significantly better fit to the Vestergaard \& Wilkes
template as shown by the upper panel of
Figure~\ref{fig:1zw1_tagn20_ncex}. While there is now a strong
predicted feature at $\lambda\,2418\,$\AA\ corresponding to
\ion{Fe}{3}, the fit to the \ion{Fe}{2} flux retains strong features for
$\lambda\le 2400\,$\AA, centred around \ion{Fe}{2}
$\lambda\,2382.04\,$\AA\ (\speco{z}{6}{F}-\spece{a}{6}{D}), which have
no counterpart (in terms of observed strength) in the empirical
template. We have also explored the possibility that the \ion{Fe}{3}
$\lambda\,2400\,$\AA\ feature is strengthened by numerous, blended
\ion{Fe}{2} transitions by merging our predicted \ion{Fe}{3} fluxes
with the more extensive \ion{Fe}{2} fluxes of SP03 based on an
829-level \ion{Fe}{2} model atom. However, such merged line-lists did
not help to improve the fit by much.

The overall fit to the \ion{Fe}{2} flux distribution is
generally better for the higher density models ($\log\,N=11.6$) but
unfortunately these models do not predict strong \ion{Fe}{3} features
for multiplet UV47, either with or without charge-exchange reactions.

\subsection{Monte Carlo Simulation Estimates for Error Bounds}

Given that all of the models presented have shortcomings when compared
to the Vestergaard \& Wilkes template, it is important to assess the
accuracy to which the iron line fluxes can be computed. This is a
complex question, encompassing everything from the accuracy of the
underlying BLR cloud model(s) and the approximations used in the
numerical methods to the accuracy of the underlying atomic data. In
this work, we will address only the latter issue by asking how
accurately the iron line fluxes can be predicted given the sometimes
large uncertainties in the basic atomic data. Even this question is not
completely straightforward.  For example, a single $A_{ji}$ value
clearly affects the flux in the $j-i$ radiative transition. However,
through the coupling of the atomic level populations by the statistical
equilibrium equations, and the global coupling of the emitting volume
through the transfer of radiation, a single $A_{ji}$ value can possibly
affect the fluxes in many lines. And similar observations can be made,
in principle, for all of the remaining atomic data, including the
photoionization and recombination rates and the collisional excitation
and de-excitation rates. To address all of these interconnections in a
consistent manner, we have turned to the Monte Carlo simulation
technique used by Sigut (1996).  Each fundamental atomic parameter
($A_{ji}$ value, effective collision strength, photoionization cross
section, collisional ionization cross section, and charge-transfer
reaction rate) is assigned an uncertainty.  Given these uncertainties,
a set of atomic data is randomly realized and used to solve the
radiative transfer-statistical equilibrium problem for the iron
fluxes.  Then, a new set of atomic data is randomly realized and
the predicted fluxes for this new data found. This sequence is
repeated, and the width of the distribution of each line flux can be
taken as a measure of its uncertainly.

Table~\ref{tab:monte} lists the basic atomic data uncertainties adopted
for our collection of iron atomic data. The error assignment is kept
deliberately simple; in principle, uncertainties could be assigned on a
transition by transition basis.  The R-matrix collision strengths (cbb
rates) for low ionization states computed under the Iron Project are
the most accurate ones available. However, they also display extremely
complicated structure due to autoionizing resonances in the
near-threshold region which dominates the rate coefficient 
at $T\sim10^{4}\,$K. Therefore, uncertainties as large as a factor of $2-3$ for
individual transitions can not be ruled out. The $A_{ji}$ values (rbb
rates) are expected to be more accurate; however, significant errors
may still be present, particularly in several calculations where
relativistic effects have not been considered in an {\it ab initio\/}
manner. Such calculations are now in progress, but the computations are
about an order of magnitude more difficult that those in LS coupling
(see Nahar, 2003).

The distribution of the random scalings is also kept very simple:  a
set of uniform random deviates, $r$, is computed with the {\sc ran2}
algorithm of Press \etal (1986). Given the uncertainty assignment, $p$,
from Table~\ref{tab:monte}, a uniform set of deviates for the logarithm
of the scaling is found from the linear relation $l=a+r(b-a)$ where
$a=-\log_{10}(p)$ and $b=\log_{10}(p)$.  The actual set of scalings,
$d$, is taken as $d=10^{l}$. This procedure ensures the following
common-sense property of the scalings: for example, if a
charge-exchange rate is assumed accurate to within a factor of 100,
then one might expect the a scaling from $0.1-1.0$ to be as likely as
one from $1-10$.  However, this procedure also results in the mean
scaling being different from one, namely
$\bar{d}=(p-p^{-1})/(ln\,10\,(b-a))$; for example, the mean scaling for
$p=100$ is about $10.8$. Thus the most probable value for the rate is
not that adopted in the default (unscaled) atomic model. However, given
the uncertainty (a factor of 100 in this case), we do not consider this
deviation significant.

Figure~\ref{fig:monte} summarizes the uncertainties due to the atomic
data uncertainties in Table~\ref{tab:monte} on the predicted
\ion{Fe}{2}-\ion{Fe}{3} flux for model u13h11. This model BLR was
chosen as it gave the best fit to the overall \ion{Fe}{2}-\ion{Fe}{3}
flux levels in the Vestergaard \& Wilkes template. Charge-exchange
reactions between iron and hydrogen were included in this calculation,
although assumed uncertain to within a factor of 10
(Table~\ref{tab:monte}).  The top panel displays the minimum and
maximum predicted flux at each wavelength and the middle panel shows
the standard deviation of each predicted flux expressed as a percentage
of the average flux at that wavelength. The largest uncertainties
approach 30\% for the \ion{Fe}{3} features near 2000\,\AA. This
reflects mostly the scaling of the charge-exchange reaction rates which
strongly affects the \ion{Fe}{2}-\ion{Fe}{3} ionization balance.  Much
smaller uncertainties are predicted for the bulk of the \ion{Fe}{2}
emission; this smaller uncertainty reflects, to some extent, the large
number of individual \ion{Fe}{2} transitions contributing to each
wavelength in the broadened spectrum (see the bottom panel of
Figure~\ref{fig:monte}), each with an individually scaled
uncertainty following Table~\ref{tab:monte}. In light of this figure,
the detailed discrepancies between the I~Zw~1 template of Vestergaard
\& Wilkes and the computed model model shown in
Figure~\ref{fig:panel_1zw1} are mostly explained as uncertainties in
the underlying atomic data.

\section{Discussion}

The primary aim of this continuing project is to incorporate advanced
methods of non-LTE radiative transfer from stellar astrophysics into
emission line analysis of AGN. In the current work, we have provided
predictions for the entire low-ionization spectrum of iron, focusing on
\ion{Fe}{3}. Given the preliminary nature of the models, and the use of
only single-zone BLR cloud models, a reasonable fit to the
\ion{Fe}{2}-\ion{Fe}{3} UV template of Vestergaard \& Wilkes for I~Zw~1
was obtained.

An aspect in which our treatment is incomplete is in the
detailed treatment of the \ion{Fe}{4} atom, the dominant ionization
state in the fully ionized optically thin region of BLR models.
The atomic data for \ion{Fe}{4} are in hand, computed from the
Iron Project, and we plan to incorporate those into a further extension
of the present work on par with \ion{Fe}{2} and \ion{Fe}{3}. The
\ion{Fe}{4} lines may also contribute to UV spectra of BLR; for
example, see observations of the Orion nebula using HST (Rubin \etal
1997).

In addition, the $\lambda\,2418\,$\AA\ feature in the computed and
observed \ion{Fe}{3} spectra offers a diagnostics of atomic processes
and physical conditions and of uncertainties in atomic data.  The
combined role of these factors in determining the intensity of this
emission feature may be illustrated by considering level-specific $\rm
e^{-} + Fe^{3+}\,\rightarrow\,Fe^{2+}$ recombination.  An increase in
the recombination-cascade rates might lead to: (a) less of a role for
charge-exchange, as inferred by the non-charge-exchange models, (b)
enhancement of the $\lambda\,2418$ feature due to recombination
contribution to the \ion{Fe}{3} line, and (c) less flux in \ion{Fe}{2}
short-ward of $2400\,$\AA, consistent with observations. The present
recombination rates were derived from level-specific photoionization
cross sections of \ion{Fe}{3} computed in LS coupling (Nahar \& Pradhan
1994). Later work has shown (Pradhan, Nahar \& Zhang 2001, Nahar \&
Pradhan 2003) that near-threshold resonance complexes may affect the
recombination rates by up to several factors.  It is necessary to
consider relativistic fine structure and resolve the resonance
complexes in great detail (excited metastable state cross sections are
affected much more than than the ground state). Nahar \& Pradhan (2003)
have developed a scheme for unified calculations of electron-ion
recombination, including both the radiative and the dielectronic
recombination processes. Such level-specific calculations for total
recombination into the \ion{Fe}{3} levels may possibly resolve the
discrepancy in the observed and computed intensities of the
$\lambda\,2418\,$\AA\ feature.

It is therefore not an exaggeration to say that in spite of the present
scale of this effort, both in terms of atomic physics and radiative
transfer, considerable work is still needed to improve the models.

\acknowledgements
\noindent
We would like to thank Marianne Vestergaard for providing the UV
I~Zw~1 iron template. This work was supported by the Natural
Sciences and Engineering Research Council of Canada (TAAS) and by the
U.S. National Science Foundation and NASA (AKP \& SN).

\begin{deluxetable}{lrr}
\tablewidth{0pt}
\tablecaption{The non-LTE Fe\,{\sc i}-{\sc iv} model atom.
\label{tab:iron_atom}} 
\tablecolumns{4}
\tablehead{
\colhead{Atom} & \colhead{Number of} & \colhead{Number of rbb } \\
\colhead{} & \colhead{NLTE Levels} & \colhead{Transitions}
}
\startdata
Fe\,{\sc i}   &  77 &   185 \\
Fe\,{\sc ii}  & 285 &  3892 \\
Fe\,{\sc iii} & 581 & 10885 \\
Fe\,{\sc iv}  &   1 &    0  \\
Total         & 944 & 14962 \\
\enddata
\end{deluxetable}

\begin{deluxetable}{llll}
\tablewidth{0pt}
\tablecaption{The single-zone BLR models. \label{tab:BLR}} 
\tablecolumns{4}
\tablehead{
\colhead{Label} & \colhead{$U_{\rm ion}$} & \colhead{$n_{\rm H}$}
& \colhead{$N_{\rm H}$} \\
\colhead{} & \colhead{} & \colhead{$\rm (cm^{-3})$} &
\colhead{$\rm (cm^{-2})$}
}
\startdata
u13h96  &$10^{-1.3}$ & $10^{9.6}$ & $10^{23}$ \\
u13h11  &$10^{-1.3}$ & $10^{11.6}$ & $10^{23}$ \\
u20h96   &$10^{-2}$ & $10^{9.6}$ & $10^{23}$ \\
u20h11   &$10^{-2}$ & $10^{11.6}$ & $10^{23}$  \\
u30h96   &$10^{-3}$ & $10^{9.6}$ & $10^{23}$ \\
u30h11   &$10^{-3}$ & $10^{11.6}$& $10^{23}$  \\
\enddata
\end{deluxetable}

%
%
\begin{deluxetable}{ccrrr}
\tablewidth{0pt}
\tablecaption{Iron fluxes for the BLR models. \label{tab:iron_per_fluxes}} 
\tablecolumns{5}
\tablehead{
\colhead{Model} & \colhead{$F_{\rm tot}$} & \colhead{\%\ion{Fe}{1}}
& \colhead{\%\ion{Fe}{2}} & \colhead{\%\ion{Fe}{3}}
}
\startdata
\cutinhead{Solar abundance, $V_t=10\,\rm km/s$} \\
u30h96& $2.246+05$   &  0.3   &  96.3   &   3.4 \\
u30h11& $5.393+06$   &  2.9   &  88.8   &   8.3 \\
u20h96& $1.479+06$   &  0.2   &  94.5   &   5.4 \\
u20h11& $2.470+07$   &  3.3   &  81.9   &  14.8 \\
u13h96& $4.698+06$   &  0.2   &  94.1   &   5.7 \\
u13h11& $1.572+08$   &  1.2   &  63.1   &  35.7 \\
\cutinhead{Solar abundance, $V_t=50\,\rm km/s$} \\
u30h96& $2.936+05$   &  0.3   &  95.9   &   3.9 \\
u30h11& $1.006+07$   &  2.1   &  92.9   &   5.0 \\
u20h96& $2.366+06$   &  0.1   &  94.8   &   5.1 \\
u20h11& $6.687+07$   &  1.5   &  89.1   &   9.4 \\
u13h96& $8.840+06$   &  0.1   &  95.2   &   4.7 \\
u13h11& $4.198+08$   &  0.5   &  74.3   &  25.2 \\
\cutinhead{3\,x\,Solar abundance, $V_t=10\,\rm km/s$} \\
u30h96& $5.958+05$   &  0.3   &  96.7   &   3.0 \\
u30h11& $1.186+07$   &  2.9   &  87.0   &  10.0 \\
u20h96& $3.471+06$   &  0.2   &  94.6   &   5.2 \\
u20h11& $4.004+07$   &  4.8   &  76.8   &  18.4 \\
u13h96& $1.003+07$   &  0.3   &  93.9   &   5.9 \\
u13h11& $2.520+08$   &  1.9   &  55.1   &  43.0 \\
\enddata
\tablecomments{The total flux, $F_{\rm tot}$ in $\rm ergs\,cm^{-2}\,s^{-1}$, is for
$1500\le\,\lambda\le\,5000\,$\AA.  Columns (3) through (5) give the
percentage contribution of each iron ion to the total flux.  The models
are identified as in Table~\ref{tab:BLR}. The notation $a+b$ means
$a\,\cdot\,10^{b}$.}
\end{deluxetable}

\begin{deluxetable}{lccl}
\tablewidth{0pt}
\tablecaption{Uncertainties assigned to the iron atomic data.
\label{tab:monte}} 
\tablecolumns{4}
\tablehead{
\colhead{Type} & \colhead{Atomic Parameter} & \colhead{Uncertainty$^{\,a}$} & \colhead{Notes}
}
\startdata
rbb    &$A_{ji}$                   & $1.50$  & \underline{R}-matrix \\
rbb    &$A_{ji}$                   & $2.00$  & semi-empirical \\
rbf$\,^b$&$\sigma_{i\kappa}(\nu)$    & $1.50$  & \underline{R}-matrix \\
cbb    &$\gamma_{ij}(T_e)$       & $1.25$  & \underline{R}-matrix \\
cbb    &$\gamma_{ij}(T_e)$       & $10.0$  & Gaunt Factor approx. \\
cbf    &$\Upsilon_{i\kappa}(T_e)$  & $10.0$  & Seaton approx.  \\
cex    &$r_{i\kappa}(T_e)$         & $10.0$ & Landau-Zener approx. \\
\enddata
\tablecomments{(a) The uncertainty is denoted $p$ in the text. (b) The recombination
rate to each individual level is related to the photoionization cross section (rbf)
through the Einstein-Milne relation.}
\end{deluxetable}

%
%

\newpage

\setcounter{figure}{0}

\begin{figure}[p]
\epsscale{0.8}
\plotone{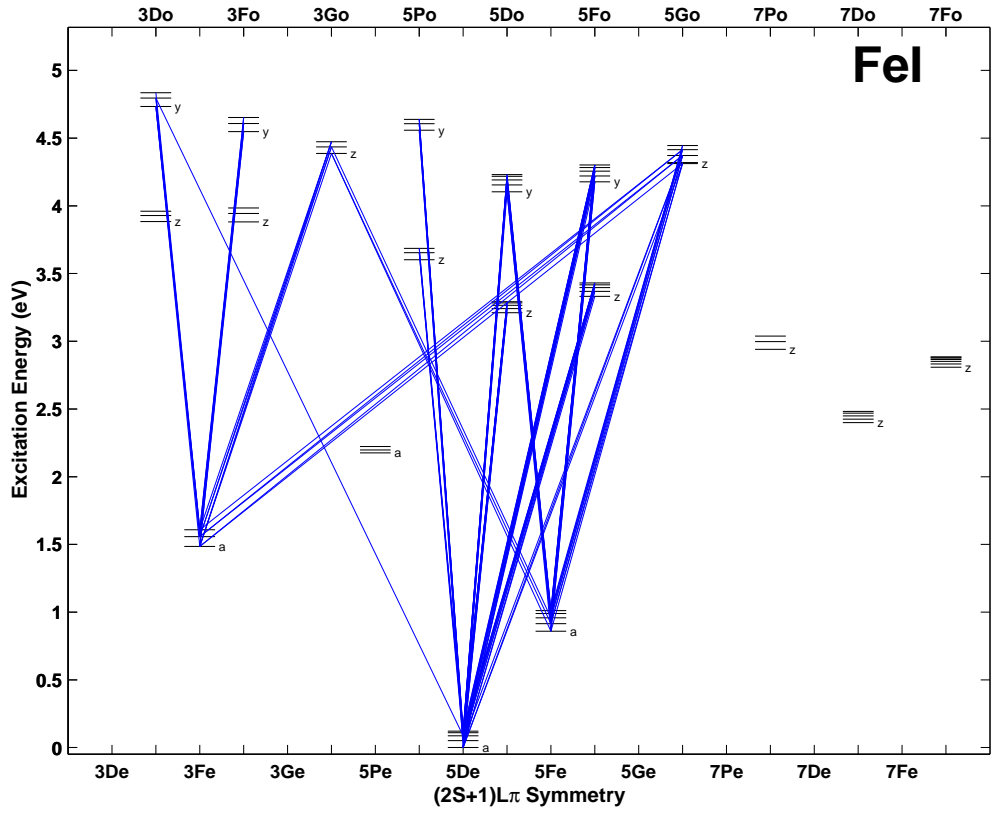}
\figcaption[f1.eps]{
A Grotrian diagram for \ion{Fe}{1} showing all of the included energy
levels and 116 ($f_{ij}>10^{-3}$ and $2500\le\lambda\le4500\,$\AA) of the
285 included radiative transitions. \label{fig:fe1_grotrian}}
\end{figure}

\begin{figure}[p]
\epsscale{0.8}
\plotone{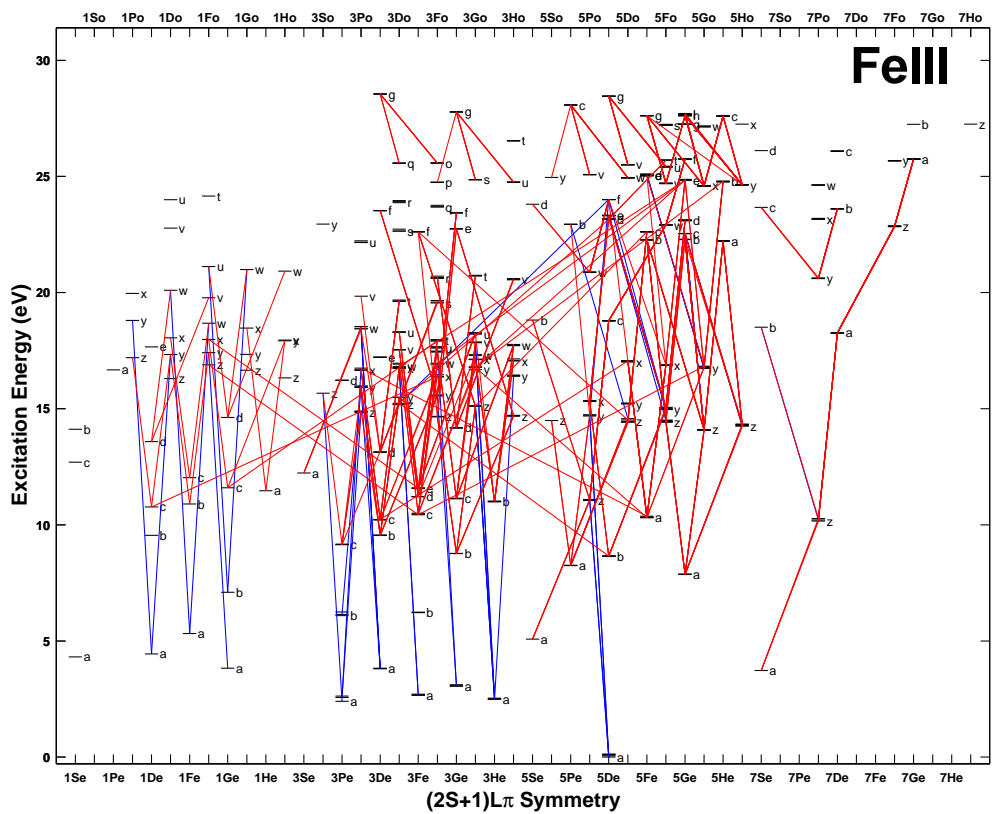}
\figcaption[f2.eps]{
A Grotrian diagram for \ion{Fe}{3} showing all of the included energy
levels and the 485 radiative transitions with $f_{ij}\,>\,0.1$
as predicted by Kurucz (1993); blue lines show transitions with
$\lambda<1500\,$\AA\ and red lines, transitions with
$1500<\lambda<5000\,$\AA.  These transitions are only a tiny fraction of
the total number of radiative transitions included in the calculation.
\label{fig:fe3_grotrian}} \end{figure}

\begin{figure}[p]
\epsscale{0.8}
\plotone{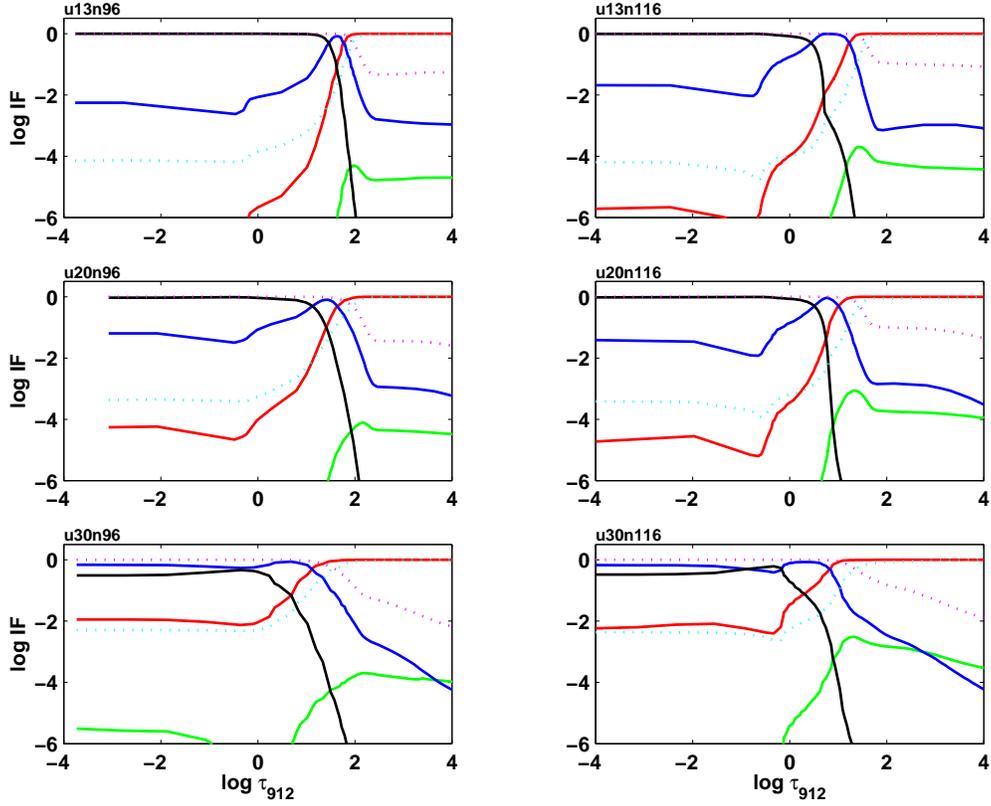}
\figcaption[f3.eps]{
Predicted \ion{Fe}{1}-\ion{Fe}{4} ionization balances for all of the
BLR models of Table~\ref{tab:BLR} assuming the solar iron abundance and
an internal turbulent velocity of $10\,\rm km\,s^{-1}$. The iron ions
can be identified as \ion{Fe}{1} (solid green line), \ion{Fe}{2} (solid
red line), \ion{Fe}{3} (solid blue line), and \ion{Fe}{4} (solid black line).  Also
shown in each panel is the hydrogen ionization balance 
decomposed into \ion{H}{1} (magenta dotted line)
and \ion{H}{2} (pink dotted line).\label{fig:ion_bal_vt10_Fep0}} 
\end{figure}

\begin{figure}[p]
\epsscale{0.8}
\plotone{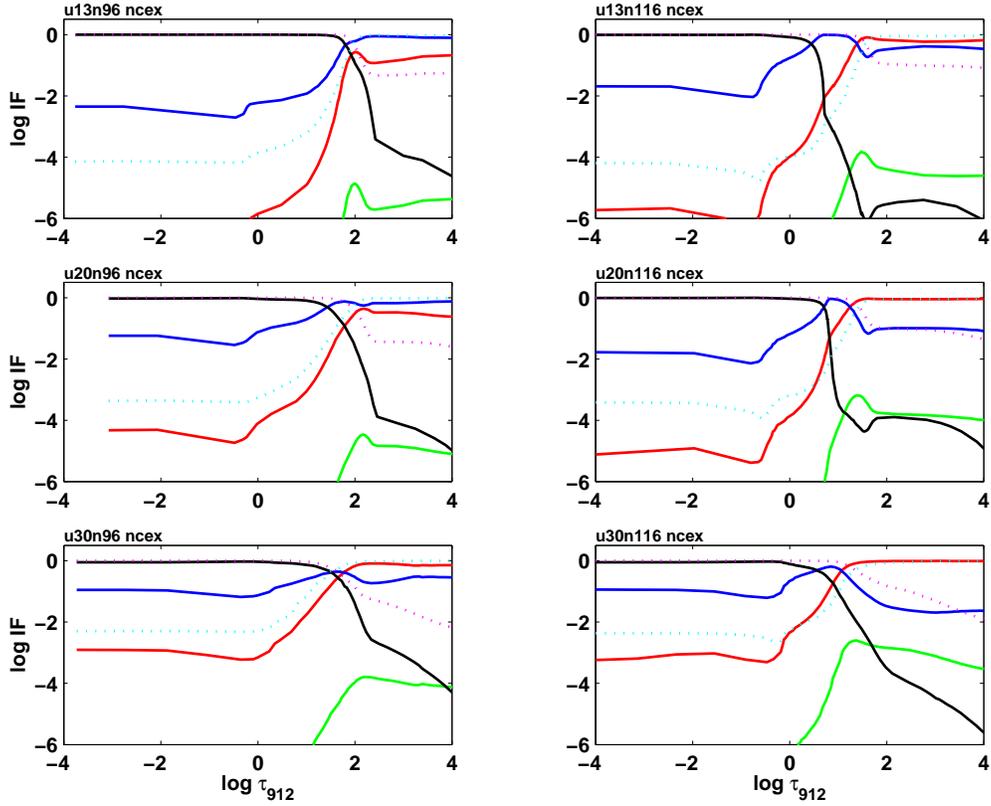}
\figcaption[f4.eps]{
Predicted \ion{Fe}{1}-\ion{Fe}{4} ionization balances in the absence of
Fe-H charge-exchange reactions for all of the
BLR models of Table~\ref{tab:BLR} assuming the solar iron abundance and
an internal turbulent velocity of $10\,\rm km\,s^{-1}$. The line styles
are the same as for Figure~\ref{fig:ion_bal_vt10_Fep0}.
\label{fig:ion_bal_vt10_Fep0_ncex}} 
\end{figure}

\begin{figure}[p]
\epsscale{0.8}
\plotone{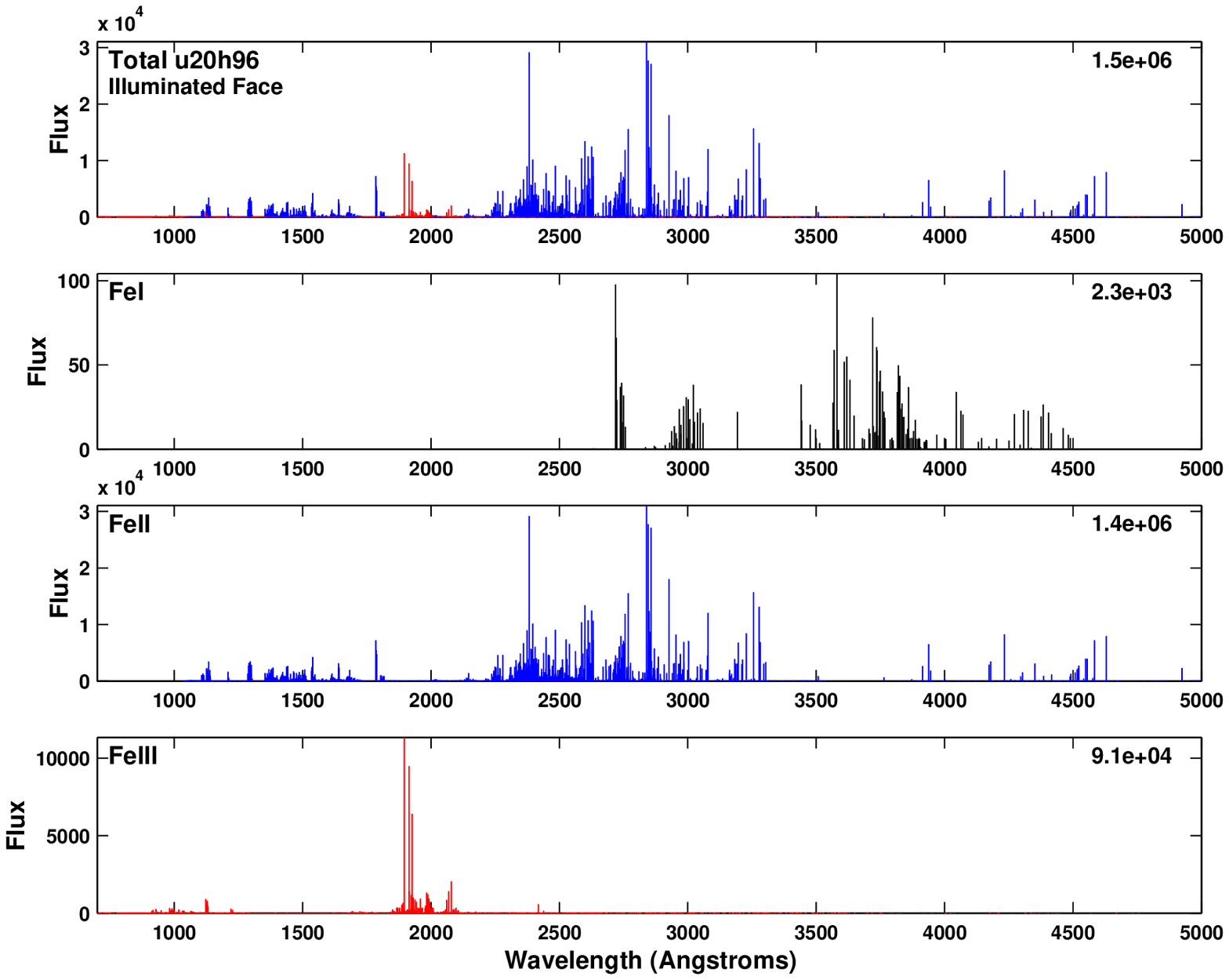}
\figcaption[f5.eps]{
The predicted iron emission spectra from the illuminated cloud face for
BLR model u20n96. The top panel gives the total \ion{Fe}{1}-\ion{Fe}{3}
line flux while the lower panels show the contributions of the
individually identified iron ions.
\label{fig:flux_tagn20_fe123}} 
\end{figure}

\begin{figure}[p]
\epsscale{0.8}
\plotone{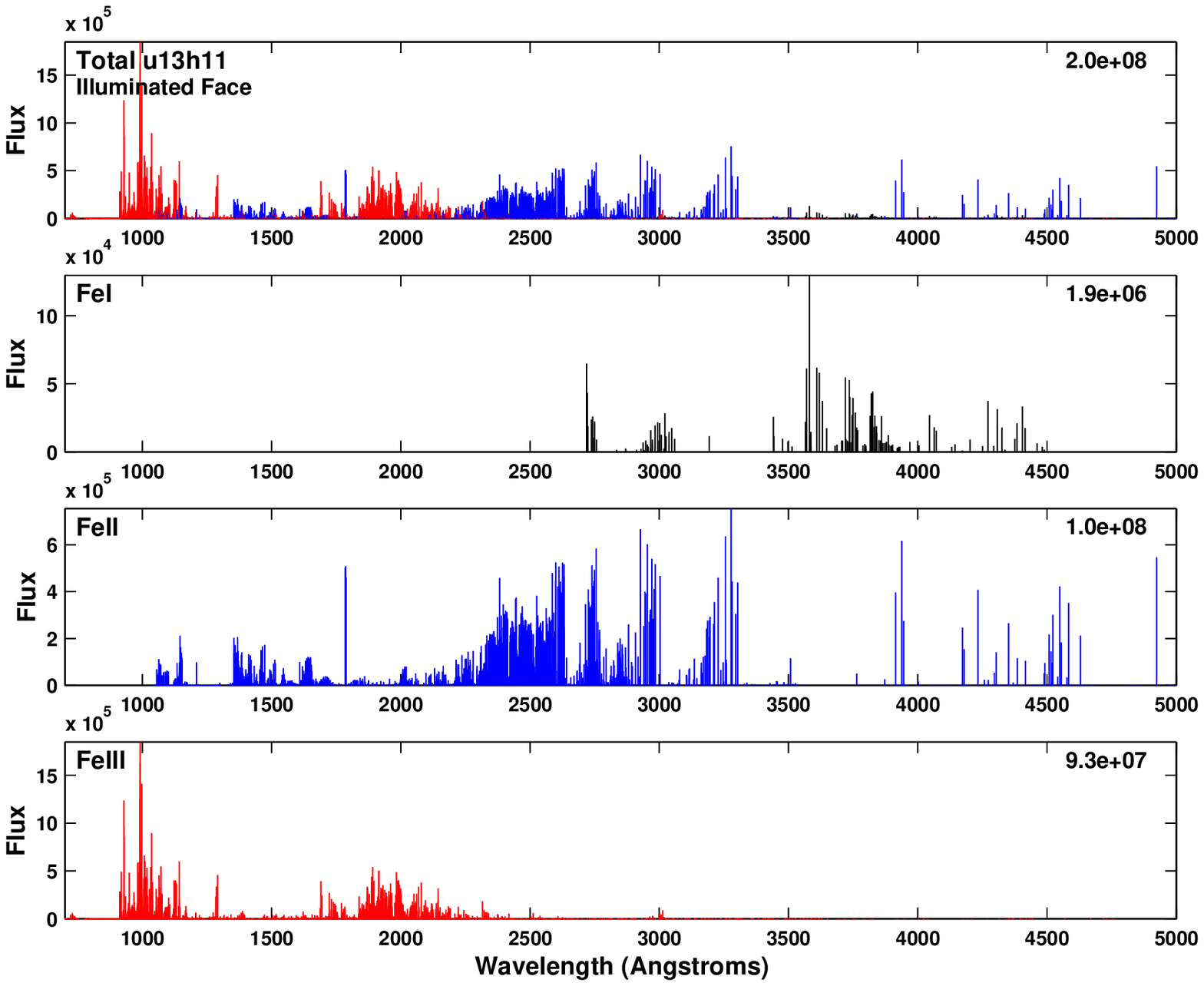}
\figcaption[f6.eps]{
The predicted iron emission spectra from the illuminated cloud face for
BLR model u13h11. The top panel gives the total \ion{Fe}{1}-\ion{Fe}{3}
line flux while the lower panels show the contributions of the
individually identified iron ions. \label{fig:flux_tagn70_fe123}} 
\end{figure}

\begin{figure}[p]
\epsscale{0.8}
\plotone{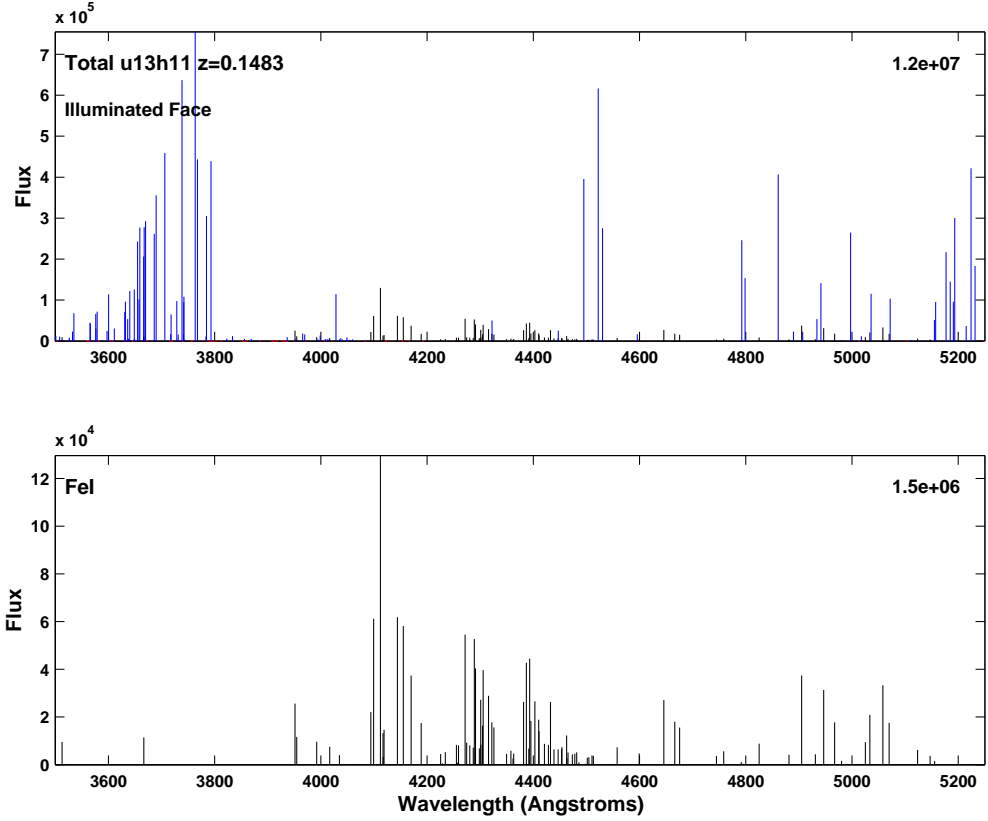}
\figcaption[f7.eps]{
The predicted iron emission spectra from the illuminated cloud face for
BLR model u13h11 in a window around the strongest \ion{Fe}{1}
transitions.  The top panel gives the total \ion{Fe}{1} (black lines)
and \ion{Fe}{2} (blue lines) flux. The lower panel shows the
\ion{Fe}{1} transitions alone.  The wavelength scale is for
the redshift of IRAS 07598+6508 ($0.1483$) so that this Figure can
be compared directly to Figure~1 of Kwan \etal (1995).
\label{fig:iras_fe1}} 
\end{figure}

\begin{figure}[p]
\epsscale{0.8}
\plotone{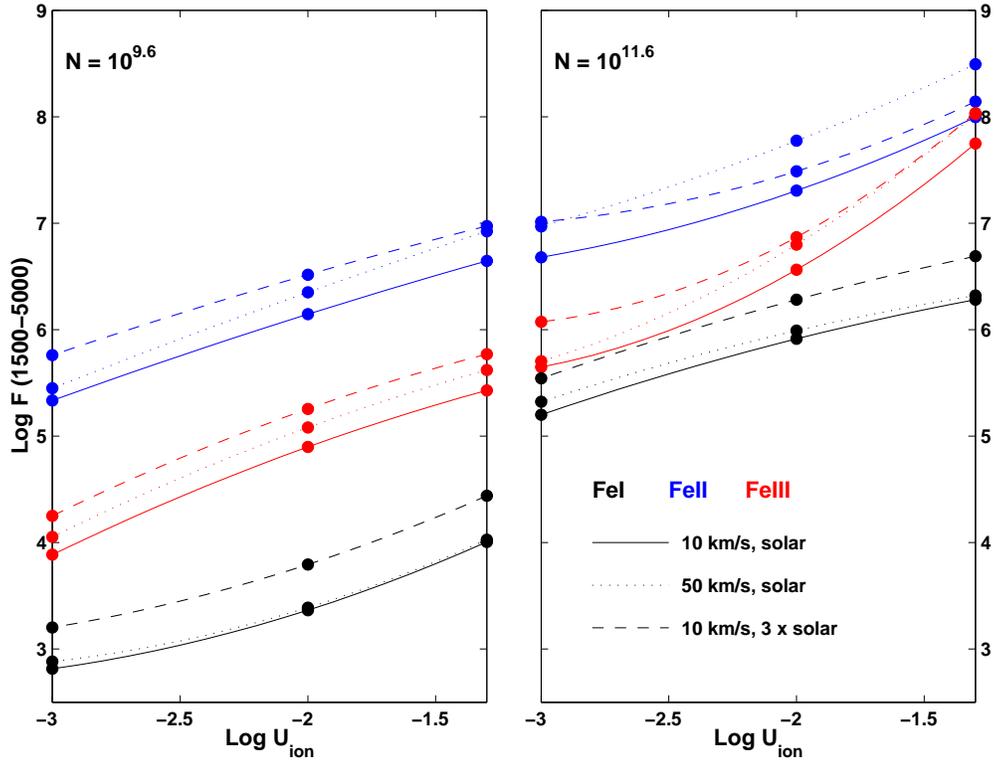}
\figcaption[f8.eps]{
Dependence of the Fe\,{\sc i}, Fe\,{\sc ii}, and Fe\,{\sc iii} line fluxes
in the wavelength interval 1500-5000\,\AA\ on the single-zone BLR cloud 
parameters of Table~\ref{tab:BLR}. The effect of tripling the iron
abundance and of increasing the cloud turbulent velocity from 10 to
50\,\kms\ is also shown. \label{fig:fe123_parplot}} \end{figure}

\begin{figure}[p]
\epsscale{0.8}
\plotone{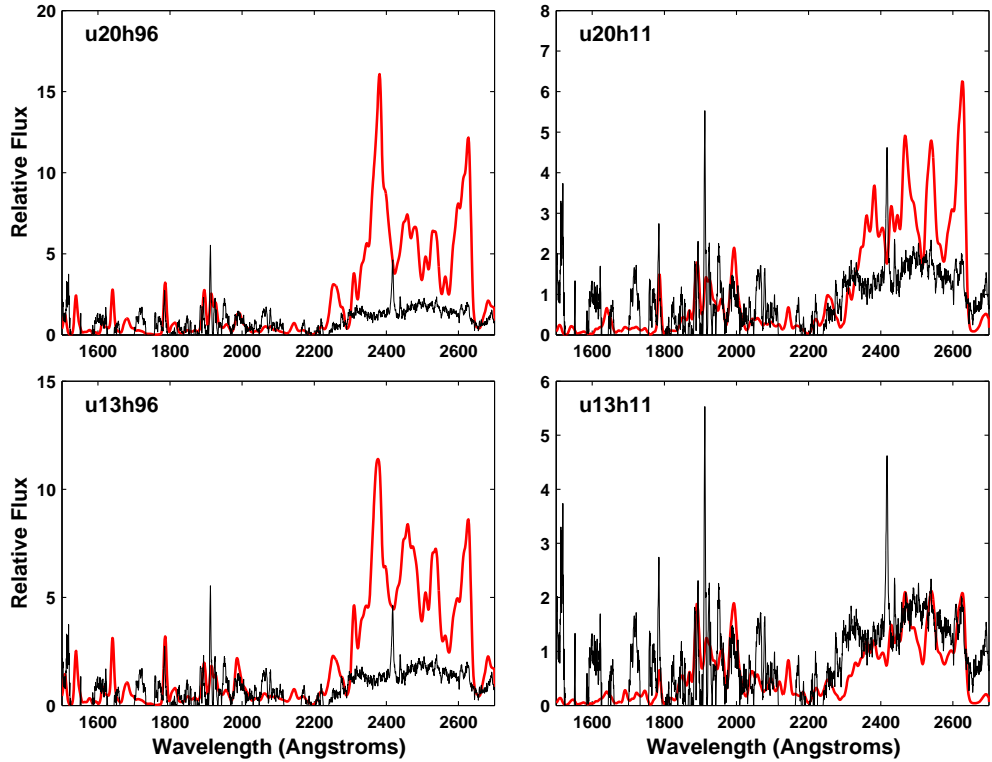}
\figcaption[f9.eps]{
A comparison of the predicted far-UV Fe\,{\sc ii}-Fe\,{\sc iii} flux
(red line) with the empirical UV Fe\,{\sc ii}-Fe\,{\sc iii} template of
Vestergaard \& Wilkes (2001, black line) for four BLR model models of
Table~\ref{tab:BLR}. All models assumed a turbulent velocity of 10\,$\rm km\,s^{-1}$
and the solar iron abundance.
\label{fig:panel_1zw1}}
\end{figure}

\begin{figure}[p]
\epsscale{0.8}
\plotone{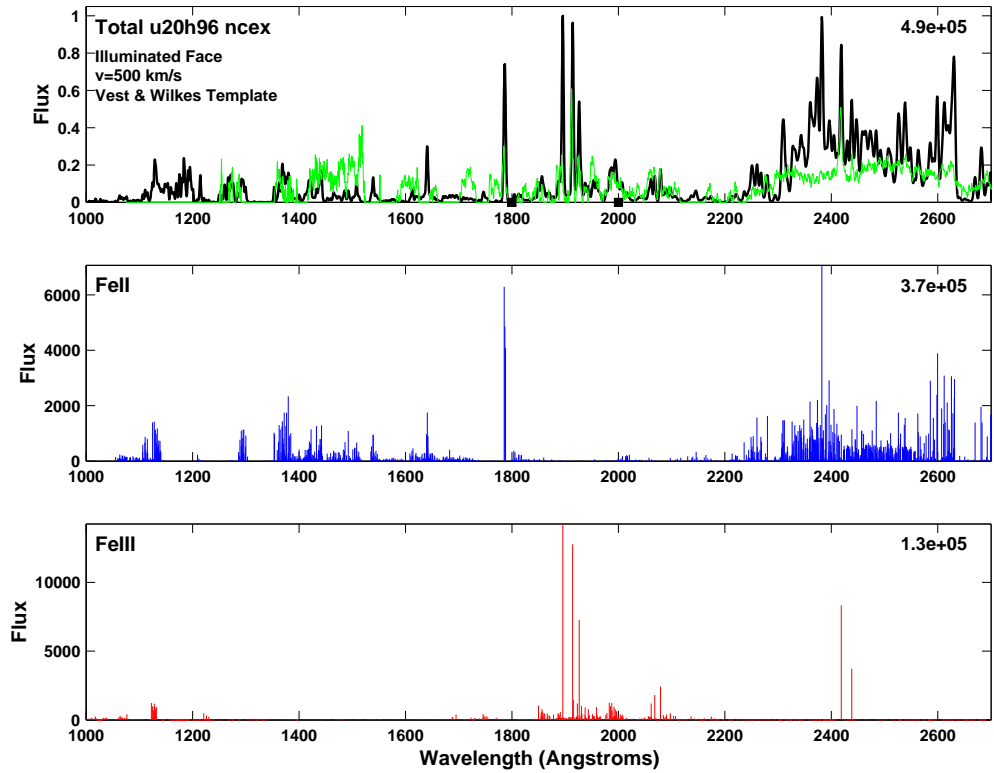}
\figcaption[f10.eps]{ 
The top panel shows a comparison of the predicted far-UV Fe\,{\sc
ii}-Fe\,{\sc iii} flux (black line) for model u20h96 without
Fe-H charge-exchange reactions with the empirical UV Fe\,{\sc
ii}-Fe\,{\sc iii} template of Vestergaard \& Wilkes (2001, green line).
The two bottom panels show the contributions of the individual iron
ions. \label{fig:1zw1_tagn20_ncex}}
\end{figure}

\begin{figure}[p]
\epsscale{0.8}
\plotone{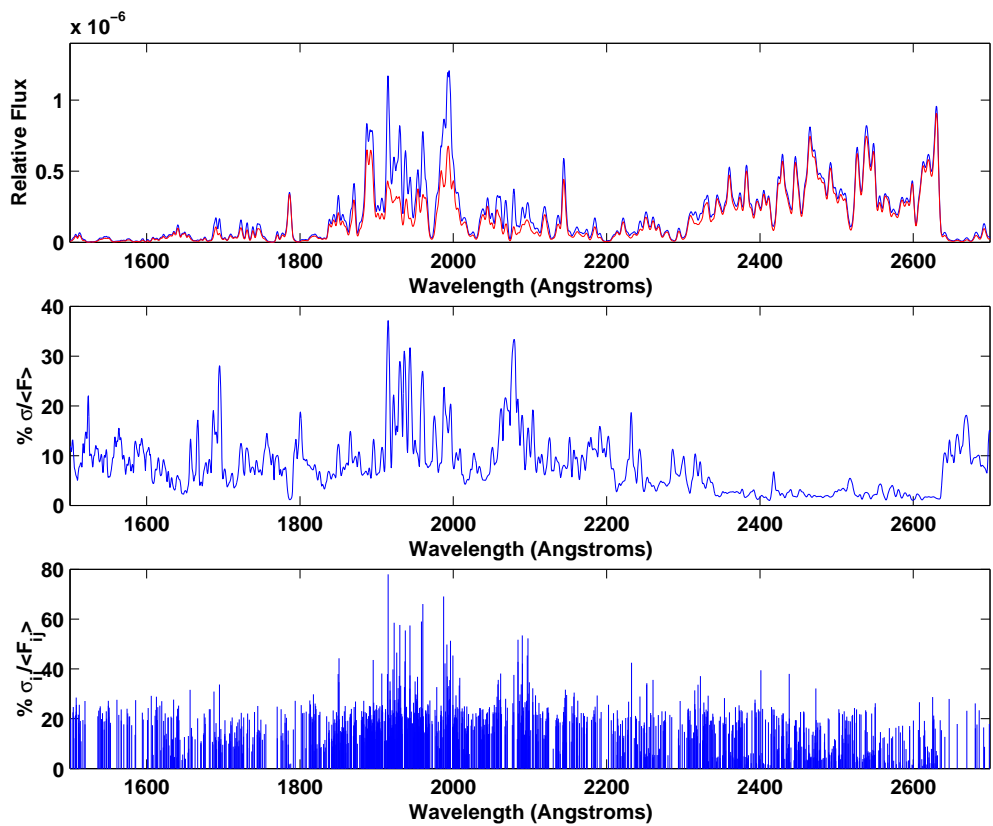}
\figcaption[f11.eps]{
The predicted UV Fe\,{\sc ii}-Fe\,{\sc iii} flux from 26 randomly
realized atomic models (see Table~\ref{tab:monte}).  The top panel
shows the minimum (red line) and maximum (read line) flux predicted at
each wavelength. The u13h11 model was used (Table~\ref{tab:BLR}) and
the spectrum was broadened with a Gaussian of FWHM of $500\,\rm
km\,s^{-1}$. The middle panel shows the standard deviation of the
predicted flux at each wavelength expressed as a percentage of the
average flux at that wavelength. The bottom panel shows the percentage
standard
deviation of all of the individually predicted iron line fluxes contributing
to the broadened spectrum.
\label{fig:monte}}
\end{figure}

\end{document}